\documentclass[aps,pra,reprint,amsmath,amssymb,superscriptaddress,nobibnotes, longbibliography]{revtex4-1}
\usepackage{graphicx,SIunits}
\usepackage{bm}     
\usepackage{hyperref} 
\usepackage{dsfont}    
\usepackage{color}

\usepackage{amsmath}
\usepackage{amsthm}
\usepackage{amssymb}
\usepackage{tikz}

\usetikzlibrary{decorations.pathreplacing}
\usepackage{braket}
\usepackage{dsfont}

\usepackage{color}
\usepackage{lineno}
\usepackage{etoolbox}

\usepackage{multirow}

\begin{document}

\preprint{APS/123-QED}

\title{Unified strategy for non-invertible Fisher information matrix in quantum metrology}

  \author{Min Namkung}%
      \affiliation{Center for Quantum Technology, Korea Institute of Science and Technology (KIST), Seoul 02792, Korea}

  \author{Changhyoup Lee}
        \email{changhyoup.lee@gmail.com}
        \affiliation{Department of Physics, Hanyang University, Seoul 04763, Korea}
        \affiliation{Hanyang Institute for Quantum Science and Quantum Technology, Hanyang University, Seoul 04763, Korea}
        \affiliation{Research Institute of Natural Sciences, Hanyang University, Seoul 04763, Korea}
  
  \author{Hyang-Tag Lim}
      \email{hyangtag.lim@kist.re.kr}
      \affiliation{Center for Quantum Technology, Korea Institute of Science and Technology (KIST), Seoul 02792, Korea}
      \affiliation{Division of Quantum Information, KIST School, Korea University of Science and Technology, Seoul 02792, Korea}

\date{\today}

\begin{abstract}
In quantum multi-parameter estimation, the precision of estimating unknown parameters is bounded by the Cram\'{e}r-Rao bound (CRB), defined via the inverse of the Fisher information matrix (FIM). However, in certain scenarios—such as distributed quantum sensing—the FIM becomes non-invertible due to parameter redundancy, which depends on the probe state and measurement. This issue is often handled using a weaker form of the CRB, potentially overestimating the uncertainty and underrepresenting achievable precision. Here, we propose an alternative approach by introducing equality constraints to remove redundancy and define the CRB via the Moore-Penrose pseudoinverse of the FIM. This framework enables systematic treatment of both simultaneous estimation and distributed sensing cases. We demonstrate its utility by reanalyzing several known examples within this unified perspective, highlighting improved interpretability and practical relevance. Our results offer a concrete guideline for addressing non-invertible FIMs and enhancing the precision of quantum multi-parameter estimation in realistic scenarios.
\end{abstract}


\maketitle

\section{Introduction}
Quantum systems are typically characterized by multiple physical parameters \cite{m.szczy,e.polino,j.liu}, making simultaneous measurement of these parameters essential across various quantum technologies. Particularly, it has attracted significant interest in quantum sensing \cite{e.roccia,e.polino2,x.guo,s.-r.zhao,l.-z.liu,d.-h.kim}, which aims to measure multiple parameters with reduced uncertainty, i.e., high precision. In estimation theory, the uncertainty in estimating the multiple parameters is quantified by the covariance matrix of the estimated parameters $\breve{ x}$, which satisfies the Cram\'{e}r-Rao inequality (CRI) \cite{h.cramer,h.l.vantrees,c.r.rao,c.w.helstrom}
\begin{equation}\label{cri_st}
    \mathrm{cov}\left(\breve{ x}\right)\succeq {F}_{ x}^{-1}.
\end{equation}
Here the lower bound is called the Cram\'{e}r-Rao bound (CRB) and is formulated in terms of the Fisher information matrix (FIM) ${F}_{ x}$ for given true parameters ${ x}$. Note that the CRB is calculated as the inverse of the FIM, which is feasible in many cases of quantum metrology \cite{l.hwang,r.demko,u.doner,m.kacprowicz,p.c.humphreys,l.pezze,c.oh,c.oh2,s.hong,j.urrehman,s.hong2,m.namkung,m.namkung2}. In particular, it is known that the FIM is less than or equal to the quantum FIM (QFIM) $F_x^{\rm (Q)}$, meaning that the quantum CRB formulated as $F_x^{\rm (Q)-1}$ lower-bounds the CRB of Eq.~(\ref{cri_st}) \cite{c.w.helstrom,p.c.humphreys,l.pezze}.

However, the FIM is not always invertible and its invertibility depends on a chosen probe state and measurement scheme that defines the FIM for a given parameter encoding scheme. Non-invertible FIMs have occasionally been observed in previous studies. For example, in the case of distributed quantum sensing \cite{z.zhang}, where the CRI of Eq.~(\ref{cri_st}) is modified by a weight vector ${w}$ under study, a weaker form of the CRB can be algebraically defined~\cite{m.gessner}. This efficiently calculable weak form has been used as the bound in the previous studies~\cite{s.-r.zhao,l.-z.liu,d.-h.kim}, rather than directly addressing non-invertible FIMs~\cite{w.ge,t.j.proctor,c.lee,l.pezze_}. However, this weak form is lower than or equal to the exact CRB \cite{m.gessner,m.malitesta}, and it may provide an optimistic bound on uncertainty. To understand the weaker form when the FIM is non-invertible, a recent work in Ref. \cite{j.wang} studied an example of a distributed quantum sensing scheme \cite{d.-h.kim}, where the inverse FIM is calculated by removing redundancy in the set of parameters being estimated. However, a rigorous parameter-reduction process still needs to be further studied such that it applies to arbitrary quantum multi-parameter estimation scenarios. Also, one could consider changing the physical setting such as the probe state \cite{x.song}, measurement scheme \cite{y.yang}, and parameter encoding~\cite{a.z.goldberg}. On the other hand, one can focus on methods to address the non-invertibility of the FIM without alternating those physical configurations in case it might not be an option in limited scenarios~\cite{a.z.goldberg,j.wang}.

In this work, we discuss a strategy using a parameter function that is used for rigorously and efficiently constraining redundant parameters, based on revisited classical parameter estimation theory~\cite{p.stoica,p.stoica0,p.stoica2,z.ben-haim,y.-h.li,a.z.goldberg}. This redundant process eventually allows an unbiased estimator to attain the CRB formulated by Moore-Penrose (MP) pseudoinverse of the FIM~\cite{r.penrose}. It consistently applies to both simultaneous estimation and distributed quantum sensing, depending on the invertibility of the FIM. We show that when the FIM is non-invertible, the key issue is to remove redundancy from the parameter set, thereby reducing it and making the FIM invertible in the reduced parameter set. We emphasize that this strategy can unify all the aforementioned methodologies without underestimating the uncertainty, and eventually apply to evaluating the QCRB from the non-invertible QFIM. Our approach further provides how to realize the effect of the projection onto FIM's support space, which has been concisely discussed in Refs.~\cite{w.ge,t.j.proctor,c.lee,l.pezze_}, by just tailoring an unbiased estimator. It thus supports that our scope is beyond calculation of the precision limit. To illustrate the proposed strategies, we investigate various cases in simultaneous estimation \cite{p.c.humphreys,l.pezze,s.hong,s.hong2,j.urrehman} and distributed quantum sensing \cite{x.guo,s.-r.zhao,l.-z.liu,d.-h.kim}. We believe our study offers simple and clear strategies applicable to any multi-parameter estimation scenario. 

\section{Proposed approach}

\subsection{Preliminaries}
Let us consider parameter-encoded states $|\psi_{ x}\rangle=\hat{U}_{ x}|\psi\rangle$ for real parameters ${x}$ and a unitary operator $\hat{U}_{ x}$. To extract information of the parameters, a measurement represented by a positive-operator-valued-measure (POVM) $\{\hat{M}_{ y}\}$ is performed on the parameter-encoded state, where $ y$ denotes measurement outcomes. From the measured outcome ${y}$, one can use an estimator that yields estimates $\breve{ x}$ of multiple parameters for given true but unknown ${x}$. Here the POVM is considered to have an element $\hat{M}_{ y}$ whose support space spans the  whole Hilbert space, such that all the parameter information is extracted from the outcome $ y$. Otherwise, some parameter may not be extracted by the POVM. The probability to obtain $ y$ is given by $p({ y}|{ x})=\langle\psi_{ x}|\hat{M}_{ y}|\psi_{ x}\rangle$ from Born's rule, and the associated FIM elements are defined by $[F]_{jk}=\sum_{{y}}\frac{1}{p({ y}|{ x})}\frac{\partial p({ y}|{ x})}{\partial y_j}\frac{\partial p({ y}|{ x})}{\partial y_k}$ \cite{h.cramer,h.l.vantrees,c.r.rao,c.w.helstrom}.

For invertible ${F}_{ x}$, the uncertainty in estimating multi-parameters ${x}$, quantified by the covariance matrix $\mathrm{cov}(\breve{{x}})=\mathbb{E}\left[(\breve{{x}}-\left\langle\breve{{x}}\right\rangle)(\breve{{x}}-\left\langle\breve{{x}}\right\rangle)^{\rm T}\right]$, for $\breve{{x}}$ being estimated parameters with an unbiased estimator $\breve{ x}={\theta}({y})$ such that $\langle\breve{{x}}\rangle={x}$ \cite{l.pezze}. In this case, an unbiased estimator yielding $\langle\breve{{x}}\rangle={x}$ exists to reach the CRB in Eq.~(\ref{cri_st}) \cite{p.stoica}. When the measurement is repetitively performed, the maximum likelihood estimator can be employed to asymptotically achieve the CRB in the limit of a large sample size \cite{l.pezze2}.

On the other hand, when the FIM is not invertible in particular cases of quantum multi-parameter estimation \cite{l.-z.liu,s.-r.zhao,d.-h.kim}, one should pay attention to using the CRB in Eq.~(\ref{cri_st}). Parameter estimation theory suggests that, in this case, the CRB is redefined as an inverse of the FIM on the support space \cite{p.stoica0,p.stoica,p.stoica2,z.ben-haim,y.-h.li}, which reads 
\begin{eqnarray}\label{cri_red}
    \mathrm{cov}(\breve{{x}})\succeq {F}_{ x}^{+},
\end{eqnarray}
where ${F}_{ x}^{+}$ denotes the MP pseudoinverse of FIM and becomes equal to ${F}_{ x}^{-1}$ when ${F}_{ x}$ is full rank, i.e., invertible. The non-invertibility of the FIM implies that there is at least one parameter which cannot be estimated with finite uncertainty. In such cases, an unbiased estimator does not exist to reach the CRB in Eq.~(\ref{cri_red}) \cite{p.stoica2}. In other words, the CRB of Eq.~(\ref{cri_red}) can only be attained after removing redundancy in the parameter set \cite{p.stoica,p.stoica0,p.stoica2,z.ben-haim,y.-h.li}. For this purpose, one can find a differentiable constraint function ${f}$ such that ${f}({x})={0}$ and $\frac{\partial {f}}{\partial{ x}^{\rm T}}{V}={0}$ \cite{y.-h.li}. Here ${V}$ is a matrix composed of orthonormal column vectors that span the support space, $\mathrm{supp}\left({F}_{  x}\right)$ \cite{p.stoica}. It follows that the gradient of ${f}$ is $\frac{\partial {f}}{\partial{  x}}=\bar{ V}$ with $\bar{ V}$ consisting of orthonormal column vectors spanning the kernel space of the FIM, $\mathrm{ker}\left({F}_{ x}\right)$. Therefore, ${f}({  x})=\bar{ V}^{\rm T}{ x}+{C}={0}$ with any constants ${C}$ \cite{y.-h.li}. Consequently, for given column vectors ${v}_j$ composing ${V}$, a parameter set is reduced to $\{{{v}}_j^{\rm T}{x}\}_{j=1}^{r}$ with $r=\mathrm{rank}({F}_{ x})$, where the FIM ${F}_{ x}$ is invertible. This process is equivalent to projection of the FIM on its support space, leading to the calculation of ${F}_{ x}^{+}$, as only concisely discussed by Refs.~\cite{w.ge,t.j.proctor,c.lee,l.pezze_}. We remark that the projection is performed by the parameter set reduction, which covers the recent study \cite{j.wang}, thereby addressing the concern in Ref.~\cite{a.z.goldberg}. It is also important to note that such reduction of the parameter set only matters to the estimation process. Therefore, it does not affect the physical setting such as the probe state, the parameter-encoding process ($\hat{U}_{ x}$), and the measurement. 

\begin{figure*}[t]
\centerline{\includegraphics[width=14cm]{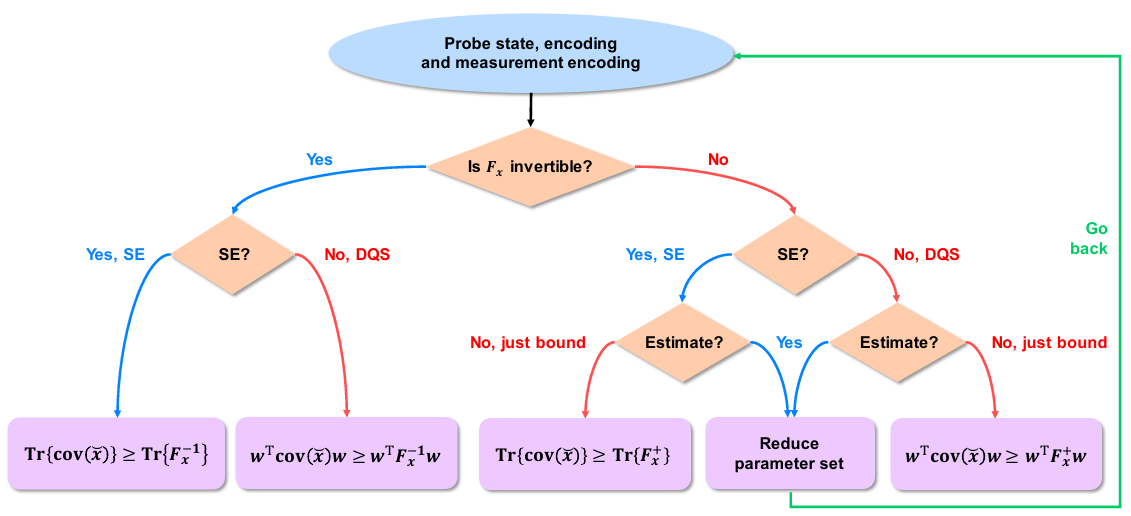}}
\caption{Flow chart for guiding how to use the CRI in multi-parameter estimation. SE and DQS denote simultaneous estimation and distributed quantum sensing, respectively.}
\centering
\label{fig1}
\end{figure*}

\subsection{Quantum sensing schemes \\ with non-invertible FIM}
We now explain how to exploit the unified CRB in Eq.~(\ref{cri_red}) in different scenarios of multi-parameter estimation. For simultaneous estimation that aims to estimate all parameters simultaneously, the total uncertainty is evaluated as $\mathrm{Tr}\left\{\mathrm{cov}(\breve{ x})\right\}$ satisfying
\begin{eqnarray}\label{cri_se}
    \mathrm{Tr}\left\{\mathrm{cov}(\breve{ x})\right\}\ge\mathrm{Tr}\left\{{F}_{ x}^{+}\right\}.
\end{eqnarray}
This applies to both invertible and non-invertible FIMs. When the FIM is invertible, ${F}_{ x}^{+}$ becomes ${F}_{ x}^{-1}$ and one can find an unbiased estimator that reaches the lower bound of Eq.~(\ref{cri_se}), eventually leading to $\mathrm{Tr}\left\{\mathrm{cov}(\breve{ x})\right\}\ge\mathrm{Tr}\left\{{F}_{ x}^{-1}\right\}$~\cite{p.c.humphreys,j.liu}. This aligns with previous studies that employed multi-mode NOON states or definite photon-number states in multiple-phase estimation \cite{p.c.humphreys,s.hong,j.urrehman,s.hong2,m.namkung,m.namkung2}. On the other hand, when the FIM is non-invertible, an unbiased estimator saturating the lower bound of Eq.~(\ref{cri_se}) does not exist. One should then remove redundancy in the set of multiple parameters by considering the aforementioned constraint function ${f}$. This process reduces the number of parameters to estimate, consequently making the FIM invertible and allowing an unbiased estimator to be found that reaches the lower bound of Eq.~(\ref{cri_se}), for the reduced parameter set. Note that the value of $\mathrm{Tr}\left\{{F}_{ x}^{+}\right\}$ for the original parameter set is equal to that of $\mathrm{Tr}\left\{{F}_{ x}'^{-1}\right\}$ with FIM ${F}_{ x}'$ for the reduced parameter set. This means that one can simply use Eq.~(\ref{cri_se}) if only the value of the lower bound is interested.

For distributed quantum sensing, where a linear combination of multiple parameters with a weight vector ${w}$ is of interest, the CRI is written as 
\begin{eqnarray}\label{cri_w}
    \left\{\Delta \left({w}^{\rm T}{ x}\right)\right\}^2\ge {w}^{\rm T}{F}_{  x}^{+}{w}.
\end{eqnarray}
This form applies to both invertible and non-invertible FIMs for any weight vector ${w}$. Here again, the invertibility of the FIM determines whether an unbiased estimator that reaches the lower bound exists. Furthermore, the parameter set can be reduced in a way that removes redundancy, making the lower bound of Eq.~(\ref{cri_w}) reachable. The CRI of Eq.~(\ref{cri_w}) is rewritten in the reduced parameter set, enabled by a projector ${\Pi}$ onto $\mathrm{supp}({F}_{ x})$, i.e., ${w}'={\Pi}{w}$ and ${F}_{ x}'={\Pi}{F}_{ x}{\Pi}$. Here, ${F}_{ x}'$ is invertible on the support space.

It is remarkable that in the literature of quantum metrology, the weaker form of the CRB of Eq.~(\ref{cri_st}) has often been used when the FIM is non-invertible for a specific weight vector ${w}$. The weak bound can also be generalized for ${F}_{ x}^+$ in this work, which is written as
\begin{eqnarray}\label{cri_weak}
    \left\{\Delta \left({w}^{\rm T}{ x}\right)\right\}^2\ge {w}^{\rm T}{F}_{  x}^{+}{w}\ge\frac{({w}^{\rm T}\sqrt{{F}_{ x}}\sqrt{{F}_{ x}}^{+}{w})^2}{{w}^{\rm T}{F}_{ x}{w}}.
\end{eqnarray} 
Here, the second inequality in Eq.~(\ref{cri_weak}) is straightforwardly proven using the Cauchy-Schwarz inequality $(\xi^{\rm T}\xi)(\zeta^{\rm T}\zeta)\ge(\zeta^{\rm T}\xi)^2$ together with $\xi=\sqrt{F_x}^{+}w$ and $\zeta=\sqrt{F_x}w$. Remarkably, the second inequality is saturated when $F_xw$ is proportional to $\sqrt{F_x}\sqrt{F_x}^+w$. This condition reduces to  $F_xw\propto w$ when the QFIM is invertible~\cite{j.wang} or when $w$ lies within $\mathrm{supp}(F_x)$ \cite{s.-r.zhao,l.-z.liu,d.-h.kim}, for both of which $\sqrt{F_x}\sqrt{F_x}^{+}$ becomes an identity. In those cases where the second inequality in Eq.~(\ref{cri_weak}) is saturated, the weak CRB provides attainable precision bound. Such a weak form has often been used to avoid directly dealing with calculations involving the non-invertible FIM. However, the weak form in Eq.~(\ref{cri_weak}) can underestimate the uncertainty when ${w}$ is not aligned with any eigenvector of the FIM, as what was studied by Ref.~\cite{m.malitesta}, regarding invertible FIMs. Note that the CRB for the exact CRB defined on the reduced parameter set coincides with the CRB given in Eq.~(\ref{cri_w}) using the pseudoinverse. Thus, {if only the bound is of interest, it suffices to evaluate Eq.~(\ref{cri_w}), which is guaranteed to be tight.} The bound of Eq.~(\ref{cri_w}) can indeed be attained using an unbiased estimator for the reduced parameter set. One can simply use our strategy for any distributed quantum sensing scenario without underestimation of the lower bound to the estimation uncertainty. To summarize, the strategies we propose are illustrated as a flowchart in Fig.~\ref{fig1}.

\subsection{Extension to QFIM}

We further remind that ${F}_{ x}$ is upper-bounded by QFIM ${F}_{ x}^{\rm (Q)}$ \cite{c.w.helstrom}, so it allows the quantum CRB to be consistently formulated with ${F}_{ x}^{\rm (Q)+}$ through Eq.~(\ref{cri_red}) to Eq.~(\ref{cri_weak}). This stresses that our methodology applies to dealing with a non-invertible QFIM. For example, when $F_{x}^{\rm (Q)}$ of a given probe state is non-invertible, $F_x$ is also non-invertible for any choice of measurements. It implies that there is at least a parameter that cannot be estimated by an unbiased estimator solely due to the structure of the probe state and encoding. Those redundant parameters can be removed according to our approach. 

In addition, when considering the QFIM, the issue of attainability should also be addressed, as it depends on the structure of the generators. First, let us assume that the generators commute with each other. In this case, optimization of the measurement enables the FIM to reach the QFIM, together with the unbiased estimator onto the reduced parameter set. Our approach then prescribes how to construct the measurement observable so as to simultaneously address both the attainability and non-invertibility issues. Conversely, even when the QFIM is not attainable due to non-commutativity~\cite{y.yang}, our method identifies which parameters cannot be estimated with finite variance under any measurements. For instance, consider a non-invertible QFIM $F_{x}^{\rm (Q)}$. It means there exists a weight $w$ such that $w^{\rm T}F_x^{\rm (Q)}w=0$. Noting that since $F_x\preceq F_x^{\rm (Q)}$, it follows directly that $w^{\rm T}F_xw=0$, implying $w\in\mathrm{ker}(F_x)$ and thus satisfying the equality constraint discussed above. In other words, there exists no measurement that can estimate $w^{\rm T}x$ with finite variance, and hence this component should be removed from the parameter set.

\section{Application to various probe states}

The above methodology has been developed and discussed in classical parameter estimation over the last few decades \cite{p.stoica,p.stoica0,p.stoica2,z.ben-haim,y.-h.li}. However, it has hardly been considered in quantum metrology, so some results already obtained with a non-invertible FIM may have been misled or misinterpreted without proper understanding of the CRB. In this work, we thus aim to apply the above methodology to quantum multi-parameter estimation: both distributed quantum sensing \cite{s.-r.zhao,x.guo,l.-z.liu} and simultaneous  multi-parameter estimation \cite{p.c.humphreys,l.pezze,s.hong,s.hong2,j.urrehman,m.namkung,m.namkung2}. In what follows, we provide simple strategies across different cases. They consequently cover typical scenarios often considered in the literature \cite{s.-r.zhao,l.-z.liu,d.-h.kim}, as will be discussed in the next section. Particularly, the constraint function described above can be applied broadly, in the strategies encapsulating the parameter transformation employed in the recent study \cite{j.wang}.

\subsection{Multiple-phase estimation}
We now apply the aforementioned strategies to paradigmatic non-trivial examples, assuming an optimized measurement setting, leading all analyses to be framed in terms of the QFIM. We investigate each scenario from the perspectives of both simultaneous estimation and distributed quantum sensing for unknown phases. Probe states that will be considered below have been employed for multiple-phase estimation scenarios \cite{x.guo,s.-r.zhao,l.-z.liu,p.c.humphreys,s.hong,j.urrehman,s.hong2,m.namkung,d.-h.kim}. 

First, consider the GHZ-like states written as
\begin{eqnarray}\label{ghz}
    |\psi_{  x }\rangle=\frac{1}{\sqrt{2}}(|v_0\rangle+e^{i{\nu}^{\rm T}{x}}|v_1\rangle),
\end{eqnarray}
which encapsulates the probe states employed in distributed quantum sensing \cite{s.-r.zhao,l.-z.liu}. It can be easily shown that the QFIM for the probe states of Eq.~(\ref{ghz}) is evaluated as
\begin{eqnarray}\label{fim_ghz}
    {F}_{  x}^{\rm (Q)}={\nu}{\nu}^{\rm T}.
\end{eqnarray}
The reason that the QFIM is non-invertible is because information of multiple parameters is concentrated to a coefficient of a single state $|v_1\rangle$. Note that the support space of the QFIM is solely spanned by the vector, implying that the QFIM for the probe states of Eq.~(\ref{ghz}) is always non-invertible. In this case, the constraint function ${f}$ is given as ${f}({x})=\bar{ V}^{\rm T}{x}+{C}$ for $\bar{{V}}$ being a matrix composed of the row vectors orthogonal to ${\nu}$ and can be used to remove redundancy in the parameter set. Thus, simultaneous estimation of the original parameter set ${x}$ is impossible through the probe state of Eq.~(\ref{ghz}). That is because an unbiased estimator can only estimate a single parameter given as ${\nu}^{\rm T}{x}$. This means that a global parameter ${w}^{\rm T}{x}$ can be estimated in distributed quantum sensing only when the weight vector ${w}\parallel{\nu}$. In addition, ${F}_{ x}^{\rm (Q)}$ in Eq.~(\ref{fim_ghz}) can be interpreted as a projector, implying that the weight vector ${w}$ is reduced to $({\nu}^{\rm T}w){\nu}$, and the redundant component $(\bar{\nu}^{\rm T}w)\bar{\nu}$ is removed from $w$ by the parameter reduction process. Notably, $({\nu}^{\rm T}w){\nu}$ is an eigenvalue of $F_x^{\rm (Q)}$, implying the weak CRB in Eq.~(\ref{cri_weak}) is attainable. We recall a probe state $|\psi_{ x}\rangle=\frac{1}{\sqrt{2}}(|HV\rangle-e^{i(x_1-2x_2)}|VH\rangle)$ studied in Ref.~\cite{s.-r.zhao}, where the support space is spanned by ${\nu}=(1,-2)$, and the weighted sum of the parameters with the weight vector ${w}=\left(1/3,-2/3\right)$ is estimated. Likewise, when $|\psi_{ x}\rangle=\frac{1}{\sqrt{2}}(|H\rangle^{\otimes 6}+e^{2i(x_1+x_2+x_3)}|V\rangle^{\otimes 6})$ is employed \cite{l.-z.liu}, a weighted sum $(x_1+x_2+x_3)/3$ is estimated in distributed quantum sensing without the reduction in these two examples, $w$ is taken to be proportional to $\nu$, making the weak CRB attainable as discussed above.

The second example considers the NOON-like probe states written as
\begin{eqnarray}\label{w}
    |\psi_{  x }\rangle&=&\frac{1}{\sqrt{m+1}}(|v_0\rangle+\sum_{j=1}^{m}e^{i\nu_j x_j}|v_j\rangle),
\end{eqnarray}
which includes the multi-mode probe states considered in simultaneous parameter estimation \cite{p.c.humphreys,s.hong,s.hong2,j.urrehman,m.namkung}. In the probe state of Eq.~(\ref{w}), each information of the parameter $x_j$ is evenly allocated to each coefficient of the state $|v_j\rangle$, leading to the invertible QFIM
\begin{eqnarray}\label{qfim_noon}
    {F}_{{  x}}^{\rm (Q)} = \frac{4\left\{(m+1)\mathrm{ diag}(\nu_1^2,\cdots,\nu_m^2)-{\nu}{\nu}^{\rm T}\right\}}{(m+1)^2}.
\end{eqnarray}
In this case, the inverse of the QFIM can be analytically evaluated by using Sherman-Morrison formula \cite{sherman}. It means that the QFIM is always invertible, and without the reduction of the parameter set, both simultaneous estimation and distributed quantum sensing can be performed with finite precision. In this case, where the QFIM is invertible, the weak CRB in Eq.~(\ref{cri_weak}) is obviously unnecessary. However, this does not imply that the weak CRB in Eq.~(\ref{cri_weak}) is always equal to the exact lower bound given by the inverse of QFIM; a counterexample can be readily found.

We lastly consider the cyclic phase-paired superposition state, taking a particular form of
\begin{eqnarray}\label{MePs}
    |\psi_{ x}\rangle=\frac{1}{\sqrt{2m}}\sum_{j=1}^{m}(|0_{j}\rangle+e^{i( x_j+ x_{j\oplus 1})}|1_{j}\rangle),
\end{eqnarray}
with $a\oplus b=a+b \ \mathrm{ mod } \ m$. This probe state encapsulates specific entangled states considered in Ref.~\cite{d.-h.kim}. In this case, the QFIM elements are calculated as
\begin{eqnarray}\label{fim_w}
    [F^{\rm (Q)}]_{jk} = \begin{cases} 
		\frac{2}{m} & \text{if }j=k, \\ 
        \frac{1}{m} & \text{if }j=k\pm1 \ \mathrm{mod} \ m, \\
        0 & \mathrm{otherwise}.
     \end{cases}
\end{eqnarray}
Here, both parameters $x_j$ and $x_{j+1}$ are encoded in a single state $|1_{j}\rangle$, rendering the QFIM non-invertible. Specifically, all eigenvalues of the QFIM are evaluated as $\lambda_k=\frac{2}{m}\left\{1+\cos\left(\frac{2\pi k}{m}\right)\right\}$ for $k\in\{1,2,\cdots,m\}$. As these eigenvalues are identical, the weak CRB in Eq.~(\ref{cri_weak}) is not tight when $w$ is the linear combination of two eigenvectors having different eigenvalues. We also observe from these eigenvalues that the presence of $\mathrm{ker}({F}_{ x}^{\rm (Q)})$ depends on $m$. For odd $m$, all the eigenvalues of the QFIM are positive, i.e. the QFIM is invertible. For even $m$, on the other hand, there is a zero eigenvalue $\lambda_{\frac{m}{2}}$, which implies that the QFIM has a one-dimensional kernel space and is therefore non-invertible. In this case, the constraint function ${f}$ is given by ${f}({x})=\bar{ V}^{\rm T}{x}+{C}$ for $\bar{ V}=[1 \ -1 \ \cdots \ 1 \ -1]^{\rm T}$, which encapsulates the recent study in Ref.~\cite{j.wang}. This can be used to reduce the original parameter set in a way that $\bar{ V}^{\rm T}{x}$ is removed, consequently making the QFIM invertible for the reduced parameter set. In the case of distributed quantum sensing, the weight vector ${w}$ also needs to be reduced to the support of the QFIM via a projector ${\Pi}={I}-\frac{\bar{ V}\bar{ V}^{\rm T}}{\left\Vert \bar{ V} \right\Vert^2}$ with an identity matrix ${I}$. It is noted that an entangled state studied by Ref.~\cite{d.-h.kim} has four modes, so there is redundancy in a parameter set, as discussed above. All  aspects of these examples are summarized in Table~\ref{table:1}.

\begin{table}[t]
    \centering
    \begin{tabular}{cccccc}
      & \multirow{2}{*}{\textbf{Eq.~(\ref{ghz})}} & \multirow{2}{*}{\textbf{Eq.~(\ref{w})}} & \multicolumn{2}{c}{\textbf{Eq.~(\ref{MePs})}}  \\ 
      &&& even $m$ & odd $m$ \\
      \hline \hline 
      Is QFIM invertible? & $\mathsf{X}$ & $\mathsf{O}$ & $\mathsf{X}$ & $\mathsf{O}$  \\
      Is weak CRB always tight? & $\mathsf{O}$ & $\mathsf{X}$ & $\mathsf{X}$ & $\mathsf{X}$  
    \end{tabular}
    \caption{Summarized properties of probe states of Eqs.~(\ref{ghz}), (\ref{w}), and (\ref{MePs}). Here, $m$ denotes the number of modes. We note that Eqs.~(\ref{ghz}), (\ref{w}), and (\ref{MePs}) encapsulate examples of Refs.~\cite{l.-z.liu,s.-r.zhao}, Refs.~\cite{m.namkung,d.-h.kim}, and Ref.~\cite{d.-h.kim}, respectively.}
    \label{table:1}
\end{table}

\subsection{Quantum metrology for many-body systems}
The examples discussed above focus on estimating parameters encoded into the phase of the probe state. We emphasize that our methodology applies constantly to estimation schemes regardless of parameter type, including parameters in a many-body system~\cite{mihailescu1,mihailescu2}. In Ref.~\cite{mihailescu1}, interaction among $N$ probe states is described by a Hamiltonian
\begin{eqnarray}
    \hat{H}=\omega\sum_{j=1}^{N}\hat{\sigma}_j^{(z)}-\sum_{j\not=k}g_{jk}\hat{\sigma}_j^{(x)} \hat{\sigma}_{k}^{(x)},
\end{eqnarray}
where $\hat{\sigma}_j^{(l)}$ are Pauli-$l$ operators on the $j$-th probe state, and $\omega$ and $g_{jk}$ denote transverse field and coupling parameters, respectively. When all $g_{jk}$ are assumed to be identical, i.e., $g_{jk}=g$, the QFIM for $w$ and $g$ is derived as~\cite{mihailescu1}
\begin{eqnarray}
    F_x^{\rm (Q)}=\sum_{k}\frac{\sin^2k}{(g^2+\omega^2-2g\omega\cos k)^2}\begin{bmatrix}
       g^2 & -g\omega \\
       -g\omega & \omega^2
    \end{bmatrix},
\end{eqnarray}
with $k=\frac{\pi(2n+1)}{N}$ and $n=0,1,2,\cdots,\frac{N}{2}-1$. In this case, the constraint function is formulated as $f(x)=\bar{V}^{\rm T}x+C=0$ with $\bar{V}=[\frac{\omega}{\sqrt{\omega^2+g^2}} \ \frac{g}{\sqrt{\omega^2+g^2}}]^{\rm T}$ and $x=[\omega \ \ g]^{\rm T}$, leading to the equality $C=-\sqrt{\omega^2+g^2}$. This limits $x$ to lie on a circle of constant radius. According to our framework, the parameter $\bar{V}^{\rm T}x$ should be removed from the parameter set, thereby focusing on a parameter $V^{\rm T}x$ with $V=[\frac{g}{\sqrt{\omega^2+g^2}} \ -\frac{\omega}{\sqrt{\omega^2+g^2}}]^{\rm T}$. However, even the reduced parameter $V^{\rm T}x$ turns out to be zero for any $w$ and $g$ when both are unknown. This implies that the frequency estimation of $\omega$ is only possible when $g$ is known as discussed by Ref.~\cite{mihailescu1}.

In Ref.~\cite{mihailescu2}, the QFIM for the identical coupling parameter $\lambda$ and the identical spin anistropy $\gamma$ in the XY model with $N=3$ is written as
\begin{eqnarray}
    F_{x}^{\rm (Q)}&=&\frac{1}{[(3\gamma^2+1)\lambda^2+4h^2+4h\lambda]^2}\nonumber\\
    && \ \  \times\begin{bmatrix}
        12\gamma^2h^2 & 6\gamma h\lambda(2h+\lambda) \\
        6\gamma h\lambda(2h+\lambda) & 3\lambda^2(\lambda+2h)^2
    \end{bmatrix},
\end{eqnarray}
yielding the constraint function $f(x)=\bar{V}^{\rm T}x+C=0$ with $\bar{V}$ orthogonal to $\vec{\nabla}_{x}\Omega$ with $\Omega=\frac{\lambda\gamma}{2h+\lambda}$~\cite{mihailescu2}. Obviously, a normalized vector $V=\frac{\vec{\nabla}_x\Omega}{\left\Vert\vec{\nabla}_x\Omega\right\Vert}$ is orthogonal to $\bar{V}$, constituting a parameter $V^{\rm T}x$ that is estimable. It also suggests that the linear transformation $x\rightarrow\frac{\vec{\nabla}_x\Omega}{\left\Vert\vec{\nabla}_x\Omega\right\Vert}x$ is needed for the parameter estimation, leading to the Jacobian transformation $F_x=F_\Omega\left\Vert\vec{\nabla}_x\Omega\right\Vert^2$ with an effective quantum Fisher information $F_\Omega$~\cite{j.liu,mihailescu2}.

\section{Conclusion}
We have discussed the unified strategy to avoid the non-invertible FIM by tailoring an unbiased estimator, thereby realizing its MP pseudoinverse, equivalent to the exact inverse on the support space. Notably, the suggested strategy covers all the previous methodologies \cite{w.ge,t.j.proctor,c.lee,j.wang}, and generalizes the weak form~\cite{m.gessner}. The unified strategy suggests simple and clear strategies across different cases of simultaneous estimation and distributed quantum sensing. Especially when the FIM is non-invertible, the CRB can be saturated by an unbiased estimator in the reduced multi-parameter set, where redundancy is removed using a constraint function proposed in Ref.~\cite{y.-h.li}. This eventually prevents the underestimation that can occur from the weak form of the CRB. To demonstrate the strategy, we have revisited various cases of simultaneous estimation \cite{p.c.humphreys,l.pezze,s.hong,s.hong2,j.urrehman} and distributed quantum sensing \cite{s.-r.zhao,l.-z.liu,x.guo,d.-h.kim}. 

We believe that our strategies developed in this work is widely applicable to a number of multi-parameter estimation scenarios \cite{t.j.proctor,d.tsarev}. For instance, when estimating parameters of a qudit system interacting with a magnetic field, the QFIM is not always invertible due to the structure of the qudit state~\cite{a.candeloro}. In this case, our methodology provides a constraint function that identifies redundant parameters, thereby enabling multiparameter estimation on the reduced parameter set. This approach contrasts with the recent study by Ref.~\cite{mihailescu3}, in which an external field is applied to remove the singularity of QFIM. Moreover, our methodology can be connected to the two-parameter estimation schemes proposed by Refs.~\cite{a.candeloro,j.he,c.mukhopadhyay}, in which $n_1$ measurement repetitions are allocated for the estimation of one parameter and $n_2$ repetitions for the other. Once the region of redundant parameters is verified using the equality constraint, the scheme of Refs.~\cite{a.candeloro,j.he,c.mukhopadhyay} can be performed to lower the uncertainty around the verified region.

\section*{Acknowledgement}
We thank A. Mitra and S. Mukherjee for discussion. This work was partly supported by National Research Foundation of Korea (NRF) grant funded by the Korea government (MSIT) (RS-2022-NR068817, RS-2024-00336079, 2023M3K5A1094813), Institute for Information \& Communications Technology Planning \& Evaluation (IITP) grant funded by the Korea government (MSIT) (RS-2023-00222863, RS-2025-02292999), and the KIST research program (2E33541, 2E33571).

\end{document}